\documentclass[preprint,preprintnumbers, prd, floatfix, superscriptaddress,nofootinbib] {revtex4-1}
\usepackage{epsfig}
\usepackage{subfigure}
\usepackage{dcolumn}% Align table columns on decimal point
\usepackage{bm}% bold math
\usepackage[usenames ,dvipsnames]{xcolor}
\usepackage{slashed}
\usepackage{graphicx,color}
\usepackage[bookmarksnumbered, pdfstartview=FitH,colorlinks,urlcolor=blue, citecolor=blue,linkcolor=blue] {hyperref}
\usepackage{appendix}
\begin{document}
\title{Investigating $Z_{cs}(3985)$ and $Z_{cs}(4000)$ exotic states in  $\Lambda_b\to Z^-_{cs}p$ decays}

\author{Yao Yu}
\email{Corresponding author: yuyao@cqupt.edu.cn}
\affiliation{Chongqing University of Posts \& Telecommunications, Chongqing, 400065, China}
 \affiliation{Department of Physics and Chongqing Key Laboratory for Strongly Coupled Physics, Chongqing University, Chongqing 401331, People's Republic of China}
\author{Zhuang Xiong}
\affiliation{Chongqing University of Posts \& Telecommunications, Chongqing, 400065, China}
\author{Han Zhang}
%\email{Corresponding author: zhanghanzzu@163.com}
\email{Corresponding author: zhanghanzzu@gs.zzu.edu.cn}
\affiliation{School of Physics and Microelectronics,
Zhengzhou University, Zhengzhou, Henan 450001, China}
\author{Bai-Cian Ke}
\email{Corresponding author: baiciank@ihep.ac.cn}
\affiliation{School of Physics and Microelectronics,
Zhengzhou University, Zhengzhou, Henan 450001, China}
\author{Jia-Wei Zhang}
%\email{Corresponding author: jwzhang@cqust.edu.cn}
\affiliation{Department of Physics, Chongqing University of Science and Technology, Chongqing, 401331,
 China}
\author{Dong-Ze He}
%\email{Corresponding author: hedongze@cqupt.edu.cn}
\affiliation{Chongqing University of Posts \& Telecommunications, Chongqing, 400065, China}
\author{Rui-Yu Zhou}
%\email{Corresponding author: zhouruiyu@cqupt.edu.cn}
\affiliation{Chongqing University of Posts \& Telecommunications, Chongqing, 400065, China}
\date{\today}

%------------------------------------------------------------------------------

\begin{abstract}
  We study the $Z_{cs}(3985)$ and $Z_{cs}(4000)$ exotic states in the
  decays of $\Lambda_b$ baryons through a molecular scenario. In the final
  state interaction, the $\Lambda_b\to \Lambda_c^+ D_s^{(*)-}$ decays are
  followed by the $\Lambda_c^+ D_s^{(*)-}$ to $Z^-_{cs}p$ rescatterings via
  exchange of a $D^{(*)}$ meson. We predict a branching fraction of
  $(3.1^{+1.4}_{-2.6})\times 10^{-4}$ for $\Lambda_b\to Z^-_{cs}p$, which can
  be measured in the $\Lambda_b\to J/\psi K^{(*)-}p$ decay. This study proposes
  a new approach to test the molecular model,
  and guides future experimental searches for the $Z_{cs}(3985)$ and
  $Z_{cs}(4000)$.
\end{abstract}
%\pacs{}
%------------------------------------------------------------------------------
\maketitle
\section{Introduction}
Exploring exotic states is an intriguing frontier in hadron physics that has
seen significant progress in the past decade. A growing number of candidates
for exotic states have been experimentally observed, as discussed in recent
reviews~\cite{Jaffe:1976ig, Chen:2016qju, Guo:2017jvc, Brambilla:2019esw, Wan:2020oxt, Wang:2023ovj}.
Among these states, the charmonium-like states, which consist of a $c\bar{c}$
pair, have formed a large family since the first observation of the $X(3872)$
in 2003~\cite{Belle:2003nnu}. Recently, a $Z_{cs}(3985)$ state, as a
strangeness-flavour partner of the $Z_{c}(3900)$ state, was observed by
BESIII~\cite{BESIII:2020qkh} in 2021 with a mass of
$3982.5^{+1.8}_{-2.6}\pm2.1$~MeV, a width of $12.8^{+5.3}_{-4.4}\pm3.0$~MeV,
and spin-parity $J^{P}=1^{+}$. This experimental observation was anticipated
in theoretical models, such as the hadronic
molecular~\cite{Guo:2017jvc, Wang:2013hga, Yang:2020nrt, Lee:2008uy, Dong:2021juy, Chen:2013wca, Chen:2021erj, Chen:2022asf},
the compact tetraquark~\cite{Ebert:2005nc, Ferretti:2020ewe}, etc.
%, and the hadro-charmonium models~\cite{Ferretti:2020ewe, Chen:2013wca}, etc.
Following the observation of $Z_{cs}(3985)$, a $Z_{cs}(4000)$ state was
discovered by LHCb~\cite{LHCb:2021uow} with a mass of $4003\pm6^{+4}_{-14}$~MeV,
a width of $131\pm15\pm26$~MeV, and $J^{P}=1^{+}$.
Although LHCb claimed there is no
evidence that $Z_{cs}(4000)$ is the same state as the $Z_{cs}(3985)$
state, Refs.~\cite{Yang:2020nrt, Ortega:2021enc} discussed the possibility that
they may correspond to the same state.
In particular, Ref.~\cite{Ortega:2021enc} demonstrates that both the BESIII
and LHCb data can be fitted simultaneously treating they as
the same state.
This attracts significant attention to the molecular model, which
naturally interprets $Z_{cs}(3985)$ and $Z_{cs}(4000)$ as
two ``$C$-parity partners''\footnote{Although $|D\bar{D}_s^{*};1^{++}\rangle$
and $|D\bar{D}_s^{*};1^{+-}\rangle$ do not have the $C$-parity, their
quantum numbers are still denoted as $J^{PC}=1^{+\pm}$ here since
they are considered as the strangeness partners of $|D\bar{D}^{*};1^{++}\rangle$
and $|D\bar{D}^{*};1^{+-}\rangle$ in the molecular
model~\cite{Chen:2022asf}.}~\cite{Chen:2021erj, Chen:2022asf, Lee:2008uy, Meng:2021rdg, Han:2022fup}:
\begin{eqnarray}\label{Zcs}
Z_{cs}(3985)&=&|D\bar{D}_s^{*};1^{++}\rangle\equiv(|D\bar{D}_s^{*}\rangle_{J=1}+|D^{*}\bar{D}_s\rangle_{J=1})/\sqrt{2}\,,\nonumber\\
Z_{cs}(4000)&=&|D\bar{D}_s^{*};1^{+-}\rangle\equiv(|D\bar{D}_s^{*}\rangle_{J=1}-|D^{*}\bar{D}_s\rangle_{J=1})/\sqrt{2}\,.
\end{eqnarray}
On the other side, Ref.~\cite{MAIANI20211616} proposes the classification of
these two strange resonances as the strange constituents of two $S$-wave
tetraquark nonets within the framework of an SU(3) quark model.

%------------------------------------------------------------------------------

\begin{figure}[htbp]
\includegraphics[width=3.2in]{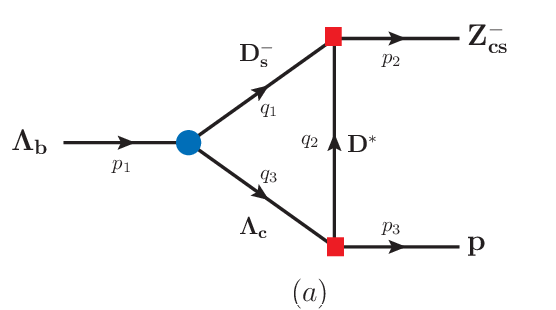}
\includegraphics[width=3.2in]{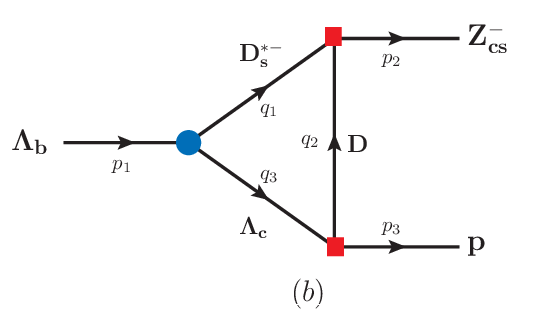}
\caption{The triangle rescattering diagrams for $\Lambda_b\to Z^-_{cs}p$
  decays.}\label{triangle1}
\end{figure}

Moreover, the production of $Z_{cs}$ can occur through the decays of $b$-flavored 
baryons. Reference~\cite{Huang:2022zsy} predicts that branching fractions of
$b$-flavored baryons two-body decays involving $Z_{cs}$ for short-distance
annihilation mechanisms are $\leq{\cal O}(10^{-7})$. However, it is also
possible for the $Z_{cs}$ states to be produced through long-distance
annihilation mechanisms, which could significantly enhance the branching
fractions~\cite{BCKa0, BCKa02}. Specifically, the long-distance annihilation
process for $\Lambda_b\to Z^-_{cs}p$ starts with the
$\Lambda_b\to \Lambda_c D_s^{(*)-}$ decay followed by $\Lambda_c$ and
$D_s^{(*)-}$ rescattering. (In the following of this report, $Z_{cs}$ denotes
$|D\bar{D}_s^{*}\rangle_{J=1}\pm|D^{*}\bar{D}_s\rangle_{J=1}$.) The
rescattering process transforms $\Lambda_c$ and $D_s^{(*)-}$ to $Z^-_{cs}$ and
$p$ via $D^{(*)}$ exchange in the triangle loop, as depicted in
Fig.~\ref{triangle1}.

In this report, we will study the $\Lambda_b\to Z^-_{cs}p$ decay with the
interpretations of $Z_{cs}$ as the molecular states and via the
triangle-rescattering diagrams. The size of the contribution from the
triangle-rescattering effect heavily depends on the couplings of the involved
intermediate interactions. Fortunately, the branching fractions of
$\Lambda_b\to \Lambda_c D_s^{(*)-}$ have been measured or estimated to be at
the $10^{-2}$ level, indicating sizable weak couplings of the baryon
decays~\cite{ParticleDataGroup:2020ssz}, and the strong couplings of
$\Lambda_c\to D^{(*)}p$ have been found to be
significant~\cite{Khodjamirian:2011sp}. In addition, the $Z_{cs}$, as
candidates for $DD_s^{*-}/D^{*}D_s^{-}$ molecular states, should strongly
couple to $DD_s^{*-}$ and $D^{*}D_s^{-}$. Therefore, the
$\Lambda_b\to Z^-_{cs}p$ decays could be dominated by long-distance effect and
study of these decays is crucial to pin down the nature of the $Z_{cs}$.
%------------------------------------------------------------------------------

%------------------------------------------------------------------------------
\section{Formalism}
With the hypothesis that the $Z_{cs}$ are $DD_s^{*-}/D^{*}D_s^{-}$ molecular
states, the triangle-rescattering processes of the $\Lambda_b\to Z^-_{cs}p$
decays, as depicted in Fig.~\ref{triangle1}, are analyzed in three parts:
$\Lambda_b\to \Lambda_cD_s^{-(\ast)}$, $\Lambda_c\to D^{*}p$ and
$Z^-_{cs}\to DD_s^{*-}(D^{*}D_s^{-})$.

The first part is the $\Lambda_b\to \Lambda_cD_s^{(*)-}$ decays, whose
amplitudes are derived to be~\cite{Cheng:1996cs,Chua:2019yqh}
\begin{eqnarray}\label{weakamp1}
{\cal M}_{1a}&\equiv&{\cal M}(\Lambda_b\to \Lambda_c D_s^{-})
=\bar{u}_{\Lambda_c}(A+B\gamma_5)u_{\Lambda_b}\,,\nonumber\\
{\cal M}_{1b}&\equiv&{\cal M}(\Lambda_b\to \Lambda_c D_s^{*-})
=\bar{u}_{\Lambda_c}(A_1\gamma_\mu\gamma_5+A_2 p_{\Lambda_c\mu}\gamma_5+B_1\gamma_\mu+B_2p_{\Lambda_c\mu})u_{\Lambda_b}\epsilon^{\mu}_{D_s^{\ast}}\,,
\end{eqnarray}
with
\begin{eqnarray}
  A &=& \frac{G_f}{\sqrt{2}}V_{cb}V^{*}_{cs}a_1 f_{D_s}(m_{\Lambda_b}-m_{\Lambda_c})(f_1+\frac{m_{\Lambda_b}^2}{m_{\Lambda_b}^2-m^2_{\Lambda_c}}f_3)\,,\nonumber\\
  B &=& \frac{G_f}{\sqrt{2}}V_{cb}V^{*}_{cs}a_1 f_{D_s}(m_{\Lambda_b}+m_{\Lambda_c})(g_1+\frac{m_{\Lambda_b}^2}{m_{\Lambda_b}^2-m^2_{\Lambda_c}}g_3)\,,\nonumber\\
  A_1 &=& -\frac{G_f}{\sqrt{2}}V_{cb}V^{*}_{cs}a_1 f_{D^*_s}m_{D^*_s}(g_1+g_2) \,,\nonumber\\
  A_2 &=& -2\frac{G_f}{\sqrt{2}}V_{cb}V^{*}_{cs}a_1 f_{D^*_s}m_{D^*_s}\frac{g_2}{m_{\Lambda_b}-m_{\Lambda_c}}\,,\nonumber\\
   B_1 &=& \frac{G_f}{\sqrt{2}}V_{cb}V^{*}_{cs}a_1 f_{D^*_s}m_{D^*_s}(f_1-f_2) \,,\nonumber\\
 B_2 &=& 2\frac{G_f}{\sqrt{2}}V_{cb}V^{*}_{cs}a_1 f_{D^*_s}m_{D^*_s}\frac{f_2}{m_{\Lambda_b}+m_{\Lambda_c}}\,,
\end{eqnarray}
%------------------------------------------------------------------------------
where $G_F$ is the Fermi constant, $V_{ij}$ the Cabibbo-Kobayashi-Maskawa
matrix elements, $f_{D^{(*)}_s}$ the decay constant, and $f_{1,3}(g_{1,3})$ the
$\Lambda_b\to \Lambda_c^+$ transition form factors, while $a_1$ results from
the factorization. The second part is the amplitude of
$\Lambda_c\to pD^{(*)}$, given by~\cite{Khodjamirian:2011sp}
\begin{eqnarray}\label{strong}
{\cal M}_{2a}&\equiv&{\cal M}(\Lambda_c\to  pD^*)=\bar{u}_p[(g^V-g^T)\gamma_\mu+2\frac{g^T}{m_{\Lambda_c}+m_p}p_\mu]u_{\Lambda_c}\epsilon^{*\mu}\,,\nonumber\\
{\cal M}_{2b}&\equiv&{\cal M}(\Lambda_c\to  pD)=ig_{\Lambda_cpD}\bar{u}_p\gamma_5u_{\Lambda_c}\,,
\end{eqnarray}
where $p_\mu$ is the momentum of $p$. The same definitions and notation as in
Ref.~\cite{Khodjamirian:2011sp} are used for the coupling constants
($g^V,g^T,g_{\Lambda_cpD}$) and the polarization four-vector
$\epsilon_\mu$. The third part is the amplitude of
$Z^-_{cs}\to DD_s^{*-}(D^{*}D_s^{-})$, written as~\cite{Wan:2020oxt}
\begin{eqnarray}
  {\cal M}_{3a}\equiv{\cal M}( Z^-_{cs}\to D^{*}D_s^-) &=&-i g_{ZD^{*}D_s}\epsilon_{Z_{cs}}\cdot\epsilon_{D^{*}}\,,\nonumber\\
  {\cal M}_{3b}\equiv{\cal M}( Z^-_{cs}\to DD_s^{*-}) &=&-i g_{ZDD_s^{*}}\epsilon_{Z_{cs}}\cdot\epsilon_{D_s^{*}}\,,
\end{eqnarray}
where $g_{ZD^{*}D_s}$ and $g_{ZDD_s^{*}}$ are the coupling constants, and
$\epsilon_{Z_{cs}}$ the polarization four-vector.
%------------------------------------------------------------------------------
Eventually, the amplitude of the triangle-rescattering processes for the
$\Lambda_b\to Z^-_{cs}p$ decay are given by
\begin{eqnarray}\label{w1}
  {\cal M}(\Lambda_b\to Z_{cs}(3985)^-p)&=&{\cal M}_{a}+e^{i\chi}{\cal M}_{b}\,,\nonumber\\
  {\cal M}(\Lambda_b\to Z_{cs}(4000)^-p)&=&{\cal M}_{a}-e^{i\chi}{\cal M}_{b}\,,
\end{eqnarray}
with
\begin{eqnarray}\label{w1p}
{\cal M}_{a(b)}&=&\int \frac{d^4{q}_{1}}{(2\pi)^{4}}
\frac{{\cal M}_{1a(b)} {\cal M}_{2a(b)} {\cal M}_{3a(b)} F_{a(b)}(q_{2}^2)}
{(q_{1}^{2}-m_{D_s(D_s^{*})}^{2})(q_{2}^{2}-m_{D^{*}(D)}^{2})(q_{3}^{2}-m_{\Lambda_c}^{2})}\,,
\end{eqnarray}
where $q_2=p_{Z_{cs}}-q_1$ and $q_3=p_{\Lambda_b}-q_1$ correspond to the
momentum flows in Fig.~\ref{triangle1}; $\chi$ represents the phase difference
between ${\cal M}_{a}$ and ${\cal M}_{b}$. The monopole form factors
$F_{a(b)}(q_{2}^2)\equiv(\lambda_{a(b)}^{2}-m_{D^{*}(D)}^{2})/(\lambda_{a(b)}^{2}-q^{2}_{2})$
with the cutoff parameters $\lambda_{a(b)}$ are to avoid the overestimation
with $q_2$ to $\pm\infty$.

%------------------------------------------------------------------------------
Since $Z_{cs}$ has $J^P=1^+$, the amplitude ${\cal M}_{a(b)}$ can be expressed
in the general form~\cite{Cheng:1996cs}:
\begin{eqnarray}\label{w2}
{\cal M}_{a(b)}=\bar{u}_{p}[{\cal A}^\prime_{1a(b)}\gamma_\mu\gamma_5+{\cal A}^\prime_{2a(b)} p_\mu\gamma_5+{\cal B}^\prime_{1a(b)}\gamma_\mu+{\cal B}^\prime_{2a(b)}p_\mu]u_{\Lambda_b}\epsilon^{\mu}_{Z_{cs}}\,.
\end{eqnarray}
To obtain ${\cal A}^\prime_{1a(b)}$, ${\cal A}^\prime_{2a(b)}$,
${\cal B}^\prime_{1a(b)}$, and ${\cal B}^\prime_{2a(b)}$, one needs to
integrate over the phase space of the triangle loop in Eq.~(\ref{w1p}). For
convenience, we define
${\cal A}^\prime({\cal B}^\prime)\equiv{\cal A}({\cal B})-\tilde{\cal A}(\tilde{\cal B})$
and factorize Eq.~(\ref{w1p}) as
\begin{eqnarray}\label{w1pp}
{\cal M}_{a(b)}=&-&\int \frac{d^4{q}_{1}}{(2\pi)^{4}}
\frac{{\cal M}_{1a(b)} {\cal M}_{2a(b)} {\cal M}_{3a(b)} }
{(q_{1}^{2}-m_{D_s(D_s^{*})}^{2})(q_{2}^{2}-m_{D^{*}(D)}^{2})(q_{3}^{2}-m_{\Lambda_c}^{2})}\nonumber\\
                &+&\int \frac{d^4{q}_{1}}{(2\pi)^{4}}
\frac{{\cal M}_{1a(b)} {\cal M}_{2a(b)} {\cal M}_{3a(b)} }
{(q_{1}^{2}-m_{D_s(D_s^{*})}^{2})(q_{2}^{2}-\lambda_{a(b)}^{2})(q_{3}^{2}-m_{\Lambda_c}^{2})}\,,
\end{eqnarray}
using
$(q_{2}^{2}-m_{D^{*}(D)}^{2})^{-1}F_{a(b)}=-(q_{2}^{2}-m_{D^{*}(D)}^{2})^{-1}+(q_{2}^{2}-\lambda_{a(b)}^{2})^{-1}$.
The first term of Eq.~(\ref{w1pp}) corresponds to ${\cal A}$ and ${\cal B}$, and
the second term $\tilde{\cal A}$ and $\tilde{\cal B}$. The integrations of the
multi-point functions are discussed in App.~\ref{point} and
Refs.~\cite{tHooft:1978jhc,Passarino:1978jh,Hahn:1998yk,Denner:2005nn}.
By comparing Eqs.~(\ref{w2}) and (\ref{w1pp}), and with help of
$\bar{u}_{p}\slashed{p}_{p}=m_p\bar{u}_{p}$,
$\slashed{p}_{\Lambda_b}u_{\Lambda_b}=m_{\Lambda_b}u_{\Lambda_b}$, and
$\sum\epsilon_{D^*(D^*_{s})}^{\mu}\epsilon_{D^*(D^*_{s})}^{*\nu}=-g^{\mu\nu}+\frac{q_{D^*(D^*_{s})}^{\mu}q_{D^*(D^*_{s})}^{\nu}}{q_{D^*(D^*_{s})}^2}$,
${\cal A}$ and ${\cal B}$ are obtained to be
\begin{eqnarray}\label{c10}
  {\cal A}_{1a}&=& \frac{g_{ZD_s D^{*}}}{16\pi^2}B[(g^V-g^T)\alpha_{11}+2\frac{g^T}{m_{\Lambda_c}+m_p}\alpha_{21}]\,,\nonumber\\
   {\cal A}_{2a}&=& \frac{g_{ZD_s D^{*}}}{16\pi^2}B[(g^V-g^T)\alpha_{12}+2\frac{g^T}{m_{\Lambda_c}+m_p}\alpha_{22}]\,,\nonumber\\
  {\cal B}_{1a}&=& \frac{g_{ZD_s D^{*}}}{16\pi^2}B[(g^V-g^T)\alpha_{13}+2\frac{g^T}{m_{\Lambda_c}+m_p}\alpha_{23}]\,,\nonumber\\
   {\cal B}_{2a}&=& \frac{g_{ZD_s D^{*}}}{16\pi^2}B[(g^V-g^T)\alpha_{14}+2\frac{g^T}{m_{\Lambda_c}+m_p}\alpha_{24}]\,,\nonumber\\
  {\cal A}_{1b}&=& \frac{g_{ZD_s^{*} D}}{16\pi^2}g_{\Lambda_cpD}[B_1\beta_{11}+B_2\beta_{21}]\,,\nonumber\\
   {\cal A}_{2b}&=& \frac{g_{ZD_s^{*} D}}{16\pi^2}g_{\Lambda_cpD}[B_1\beta_{12}+B_2\beta_{22}]\,,\nonumber\\
  {\cal B}_{1b}&=& \frac{g_{ZD_s^{*} D}}{16\pi^2}g_{\Lambda_cpD}[B_1\beta_{13}+A_2\beta_{23}]\,,\nonumber\\
   {\cal B}_{2b}&=& \frac{g_{ZD_s^{*} D}}{16\pi^2}g_{\Lambda_cpD}[B_1\beta_{14}+B_2\beta_{24}+A_1\beta_{34}+A_2\beta_{44}]\,,
\end{eqnarray}
with 
\begin{eqnarray}
  \alpha_{11}&=&m_p(C_0+C_{12}+C_{24})-m_{\Lambda_c}( C_0-C_{24})+m_{\Lambda_b}C_{11}\,,\nonumber\\
  \alpha_{12}&=&-2(C_0+C_{12})+(m_p+m_{\Lambda_c})(-m_{\Lambda_b}D_{21}+m_pD_{22}+(m_p-m_{\Lambda_b})D_{23})+m_{\Lambda_b}^2D_{31}\,,\nonumber\\
  &&-m_p^2D_{32}-(m_{\Lambda_b}^2-m_{Z_{cs}}^2+2m_p^2)D_{33}-(2m_{\Lambda_b}^2-m_{Z_{cs}}^2+m_p^2)D_{34}+6D_{35}+6D_{36}\,,\nonumber\\
  \alpha_{21}&=& -\frac{1}{2}(m_{\Lambda_b}^2-m_{Z_{cs}}^2+m_p^2)D_{35}-m_p^2D_{36}\,,\nonumber\\
  \alpha_{22}&=&-m_p(C_0+C_{12}+D_{24}+D_{35}+2D_{36})-m_{\Lambda_c}(C_0+D_{24})+m_{\Lambda_b}(C_{11}+D_{35})\,,\nonumber\\
  &&+\frac{1}{2}(m_{\Lambda_b}^2-m_{Z_{cs}}^2)[m_{\Lambda_b}(D_{31}+D_{34})-m_{\Lambda_c}(D_{21}+D_{23})-m_p(D_{21}+D_{23}+D_{33}+D_{34})]\,,\nonumber\\
  &&-\frac{1}{2}m_p^2[m_{\Lambda_c}(D_{21}+2D_{22}+3D_{23})-m_{\Lambda_b}(D_{31}+2D_{33}+3D_{34})]\,,\nonumber\\
  &&-\frac{1}{2}m_p^3(D_{21}+2D_{22}+3D_{23}+2D_{32}+3D_{33}+D_{34})\,,\nonumber\\
  \alpha_{31}&=&m_p(C_0+C_{12}+C_{24})-m_{\Lambda_c}( C_0-C_{24})-m_{\Lambda_b}C_{11}\,,\nonumber\\
  \alpha_{32}&=&-2(C_0+C_{12})-(m_p+m_{\Lambda_c})(-m_{\Lambda_b}D_{21}+m_pD_{22}+(m_p-m_{\Lambda_b})D_{23})+m_{\Lambda_b}^2D_{31}\,,\nonumber\\
  &&-m_3^2D_{32}-(m_{\Lambda_b}^2-m_{Z_{cs}}^2+2m_p^2)D_{33}-(2m_{\Lambda_b}^2-m_{Z_{cs}}^2+m_p^2)D_{34}+6D_{35}+6D_{36}\,,\nonumber\\
  \alpha_{41}&=& -\frac{1}{2}(m_{\Lambda_b}^2-m_{Z_{cs}}^2+m_p^2)D_{35}-m_p^2D_{36}\,,\nonumber\\
  \alpha_{42}&=&-m_p(C_0+C_{12}+D_{24}+D_{35}+2D_{36})-m_{\Lambda_c}(C_0+D_{24})-m_{\Lambda_b}(C_{11}+D_{35})\,,\nonumber\\
  &&+\frac{1}{2}(m_{\Lambda_b}^2-m_{Z_{cs}}^2)[-m_{\Lambda_b}(D_{31}+D_{34})-m_{\Lambda_c}(D_{21}+D_{23})-m_p(D_{21}+D_{23}+D_{33}+D_{34})]\,,\nonumber\\
  &&-\frac{1}{2}m_p^2[m_{\Lambda_c}(D_{21}+2D_{22}+3D_{23})+m_{\Lambda_b}(D_{31}+2D_{33}+3D_{34})]\,,\nonumber\\
  &&-\frac{1}{2}m_p^3(D_{21}+2D_{22}+3D_{23}+2D_{32}+3D_{33}+D_{34})\,,\nonumber\\
  \beta_{11}&=&m_{\Lambda_b}(-C_0+C_{11})+m_{\Lambda_c}C_0+m_pC_{12}\,,\nonumber\\
  \beta_{12}&=&2(C_0-C_{11})+m_{\Lambda_b}^2D_{31}+m_p^2D_{32}+(m_{\Lambda_b}^2+2m_p^2-m_{Z_{cs}}^2)D_{33}\,,\nonumber\\
    &&+(2m_{\Lambda_b}^2+m_p^2-m_{Z_{cs}}^2)D_{34}+6D_{35}+6D_{36}\,,\nonumber\\
 \beta_{21}&=& -\frac{1}{2}(m_{\Lambda_b}^2-m_{Z_{cs}}^2+m_p^2)D_{36}-m_{\Lambda_b}^2D_{35}\,,\nonumber\\
   \beta_{22}&=&-m_{\Lambda_b}(-C_0+C_{11})-m_{\Lambda_c}C_0+m_pC_{12}+m_{\Lambda_b}^3D_{31}-m_p\frac{m_{\Lambda_b}^2-m_{Z_{cs}}^2+m_p^2}{2}D_{32}\,,\nonumber\\
     &&-[\frac{m_{\Lambda_b}^2-m_{Z_{cs}}^2+m_p^2}{2}(m_p-m_{\Lambda_b})+m_{\Lambda_b}^2m_p]D_{33}-[\frac{m_{\Lambda_b}^2+m_{Z_{cs}}^2-m_p^2}{2}m_{\Lambda_b}+m_{\Lambda_b}^2(m_p+m_{\Lambda_b})]D_{34}\,,\nonumber\\
     &&m_{\Lambda_b}D_{35}+(m_{\Lambda_b}-m_p)D_{36}\,,\nonumber\\
      \beta_{13}&=&m_{\Lambda_b}(C_0-C_{11})+m_{\Lambda_c}C_0+m_pC_{12}\,,\nonumber\\
    \beta_{23}&=&-\frac{1}{2}(m_{\Lambda_b}^2-m_{Z_{cs}}^2+m_p^2)D_{36}-m_{\Lambda_b}^2D_{35}\,,\nonumber\\
  \beta_{14}&=& 2(C_0-C_{11})\,,\nonumber\\
    \beta_{24}&=&m_{\Lambda_b}(C_0-C_{11})-m_{\Lambda_c}C_0-m_pC_{12}\,,\nonumber\\
    \beta_{34}&=&m_{\Lambda_b}^2D_{31}+m_p^2D_{32}+(m_{\Lambda_b}^2+2m_p^2-m_{Z_{cs}}^2)D_{33}+(2m_{\Lambda_b}^2+m_p^2-m_{Z_{cs}}^2)D_{34}+6D_{35}+6D_{36}\,,\nonumber\\
    \beta_{44}&=&m_{\Lambda_b}^3D_{31}-m_p\frac{m_{\Lambda_b}^2-m_{Z_{cs}}^2+m_p^2}{2}D_{32}-[\frac{m_{\Lambda_b}^2-m_{Z_{cs}}^2+m_p^2}{2}(m_p+m_{\Lambda_b})+m_{\Lambda_b}^2m_p]D_{33}\,,\nonumber\\
     &&+[\frac{m_{\Lambda_b}^2+m_{Z_{cs}}^2-m_p^2}{2}m_{\Lambda_b}-m_{\Lambda_b}^2(m_p+m_{\Lambda_b})]D_{34}-m_{\Lambda_b}D_{35}-(m_{\Lambda_b}+m_p)D_{36}\,.
\end{eqnarray}
The definitions of $C_0$, $C_{ij}$, $D_{ij}$ are the same as in
Refs~\cite{tHooft:1978jhc,Passarino:1978jh,Hahn:1998yk,Denner:2005nn}. In the
same way, $\tilde{\cal A}(\tilde{\cal B})$ are obtained by replacing $m_{D^*}$
and $m_{D}$ in Eq.~(\ref{c10}) with $\lambda_{a(b)}$. Following
Ref.~\cite{Cheng:1996cs}, the decay width is determined to be
\begin{eqnarray}
  \Gamma(\Lambda_b\to Z^-_{cs}p) &=&\frac{ p_c(m_{p}+E_{p})}{4\pi m_{\Lambda_b}}[2|S|^2+|P_2|^2+\frac{E^2_{Z_{cs}}}{m^2_{Z_{cs}}}(|S+D|^2+|P_1|^2)]\,,
\end{eqnarray}
with
\begin{eqnarray}
  S &=& -({\cal A}^\prime_{1a}+e^{i\chi}{\cal A}^\prime_{1b})\,,\nonumber\\
  P_1 &=& -\frac{p_c}{m_{Z_{cs}}}[\frac{m_{p}+m_{\Lambda_b}}{E_{p}+m_{p}}({\cal B}_{1a}+e^{i\chi}{\cal B}_{1b})+m_{\Lambda_b}({\cal B}_{2a}+e^{i\chi}{\cal B}_{2b})]\,,\nonumber\\
  P_2 &=& \frac{ p_c}{E_{p}+m_{p}}({\cal B}_{1a}+e^{i\chi}{\cal B}_{1b})\,,\nonumber\\
  D &=&  -\frac{ p^2_c}{E_{Z_{cs}}(E_{p}+m_{p})}[({\cal A}^\prime_{1a}+e^{i\chi}{\cal A}^\prime_{1b})-m_{\Lambda_b}({\cal A}^\prime_{2a}+e^{i\chi}{\cal A}^\prime_{2b})]\,,
\end{eqnarray}
  where $p_c$ is the center-of-mass momentum.
%------------------------------------------------------------------------------

\section{Numerical results and Discussions}
\begin{figure}[htbp]
\includegraphics[width=3.2in]{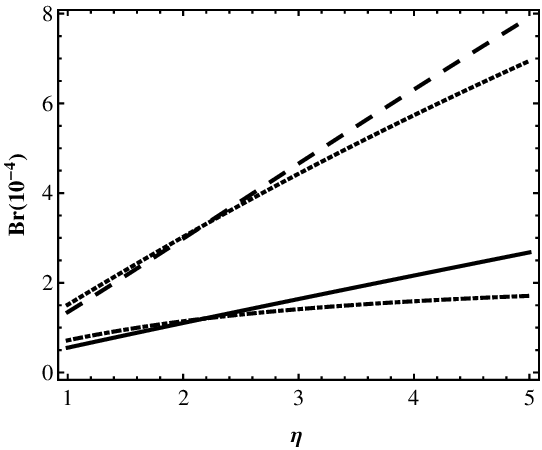}
\caption{The branching fractions of $\Lambda_b\to Z_{cs}(3985)^-p$ related to $\eta$.
  The solid, dashed, doted, and dot-dashed lines indicate branching fractions
  for $\chi=0$, $\chi=\pi/2$, $\chi=\pi$, and $\chi=-\pi/2$, respectively.}
\label{triangle2}
\end{figure}

\begin{figure}[htbp]
\includegraphics[width=3.2in]{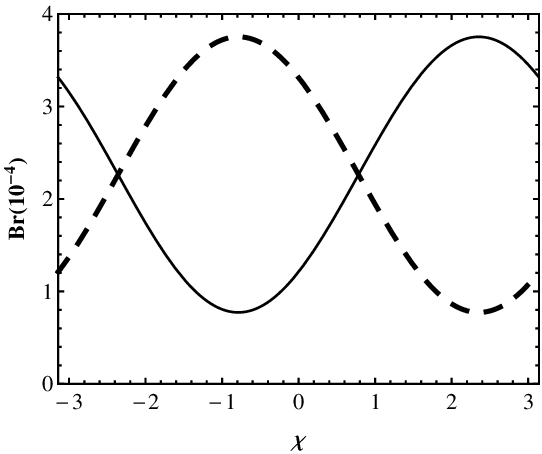}
\caption{The branching fractions of $\Lambda_b\to Z^-_{cs}p$ related to $\chi$.
  The dashed line represents $Z_{cs}(3985)^-$ and the solid line $Z_{cs}(4000)^-$.}
\label{triangle3}
\end{figure}

In the numerical analysis, we use the values of
$(V_{cb},V_{cs})=(A\lambda^2,1-\lambda^2/2)$ with $A= 0.790\pm 0.017$ and
$\lambda=0.22650\pm 0.00048$~\cite{ParticleDataGroup:2020ssz}. For the decay
constants and form factors, it is given that
$(f_{D_s},f_{D^*_s})=(0.249,0.230)$~GeV,
$(f_1,f_2,f_3)=(0.564^{+0.082}_{-0.086},-0.195^{+0.034}_{-0.037},0.089^{+0.027}_{-0.028})$,
$(g_1,g_2,g_3)=(0.555^{+0.079}_{-0.083},0.039^{+0.007}_{-0.009},-0.085^{+0.011}_{-0.011})$,
and $a_1\simeq1.0$~\cite{Chua:2019yqh}, where $a_1$ of ${\cal O}(1.0)$ is
regarded to well interpret
$\Lambda_b\to\Lambda_c D_s^-$~\cite{ParticleDataGroup:2020ssz}. We also
adopt $(g_{ZD_s D^*},g_{ZD_s^{*} D})=(3.72\pm0.78,-4.18\pm0.88)$~GeV from
Ref.~\cite{Wan:2020oxt} and
$(g^V,g^T,g_{\Lambda_cpD})=(-5.8^{+2.1}_{-2.5},3.6^{+2.9}_{-1.8},10.7^{+5.3}_{-4.3})$
from Ref.~\cite{Khodjamirian:2011sp}. The cutoff parameters are fixed to be
$(\lambda_a,\lambda_b)=(m_{D^*}+\eta\Lambda_{\rm QCD},m_{D}+\eta\Lambda_{\rm QCD})$~\cite{Cheng:2004ru}
with $\eta=2.2$ and $\Lambda_{\rm QCD}=0.22$~GeV. As a consequence, the
branching fraction of $\Lambda_b\to Z^-_{cs}p$ is determined to be
\begin{eqnarray}\label{pre_br}
{\cal B}(\Lambda_b\to Z^-_{cs}p)&=&(3.1^{+1.4}_{-2.6})\times 10^{-4}\,.
\end{eqnarray}
The uncertainty is evaluated by repeating the calculation after varying the
parameters according to their uncertainties. The phase $\chi$ is considered as
a source of the uncertainty and varied from $-\pi$ to $+\pi$. To illustrate
the dependence on $\eta$ and $\chi$, the branching fractions of
$\Lambda_b\to Z^-_{cs}p$ related to $\eta$ and $\chi$ are shown in
Figs.~\ref{triangle2} and ~\ref{triangle3}, respectively. It is interesting to
note that the $\Lambda_b\to Z_{cs}(3985)^-p$ and $\Lambda_b\to Z_{cs}(4000)^-p$
decays differ by a phase of $\pi$, which always ensures that one of them has a
branching fraction $\gtrsim 2\times10^{-4}$. 

%------------------------------------------------------------------------------

Approximately, we present the resonant branching fractions as
\begin{eqnarray}
&{\cal B}&(\Lambda_b\to Z_{cs}(3985)^-p\to J/\psi K^{*-}p)
\simeq
{\cal B}(\Lambda_b\to Z_{cs}(3985)^-p){\cal B}(Z_{cs}(3985)^-\to J/\psi K^{*-})\,,\nonumber\\
&{\cal B}&(\Lambda_b\to Z_{cs}(4000)^-p\to J/\psi K^-p)
\simeq
{\cal B}(\Lambda_b\to Z_{cs}(4000)^-p){\cal B}(Z_{cs}(4000)^-\to J/\psi K^-)\,,
\end{eqnarray}
and subsequently predict
\begin{eqnarray}
&{\cal B}&(\Lambda_b\to Z_{cs}(3985)^-p\to J/\psi K^{*-}p)
=(3.4^{+1.5}_{-2.8})\times 10^{-6}\,,\nonumber\\
&{\cal B}&(\Lambda_b\to Z_{cs}(4000)^-p\to J/\psi K^-p)
=(2.4^{+1.1}_{-2.0})\times 10^{-6}\,,
\end{eqnarray}
with $({\cal B}(Z_{cs}(3985)^-\to J/\psi K^{*-}),{\cal B}(Z_{cs}(4000)^-\to J/\psi K^-))=(1.08\%,0.78\%)$~\cite{Chen:2021erj}.

The resonance states of $\Lambda_c$ can also contribute to the
branching fractions of $\Lambda_b\to Z^-_{cs}p$.
Here, we provide a rough estimate of the contribution of the lightest resonance, $\Lambda_c(2595)$.
One could expect that
heavier resonance state will contribute less or at the same order. 
According to Ref~\cite{Chua:2019yqh}, the branching
fractions of the $\Lambda_b\to \Lambda_c D_s^{(*)-}$ are as follows
\begin{eqnarray}\label{eq1}
  {\cal B}(\Lambda_b\to \Lambda_c D_s^{-}) &=& 11.92^{+7.69}_{-5.28}\times10^{-3}\,,\nonumber\\
   {\cal B}(\Lambda_b\to \Lambda_c D_s^{*-})&=& 17.49^{+10.60}_{-7.48}\times10^{-3}\,,
\end{eqnarray}
and these of the $\Lambda_b\to \Lambda_c(2595) D_s^{-(*)}$ are given by
\begin{eqnarray}\label{eq2}
  {\cal B}(\Lambda_b\to \Lambda_c(2595) D_s^{-}) &\cong&(1.72^{+1.71}_{-1.01})\times10^{-3}\,, \nonumber\\
  {\cal B}(\Lambda_b\to \Lambda_c(2595) D_s^{*-}) &\cong&(2.28^{+2.21}_{-1.29})\times10^{-3}\,,
\end{eqnarray}
with the strong coupling of $\Lambda_c\to  pD^{(*)}$ in Eq.~(\ref{strong})
and the strong coupling of $\Lambda_c(2595)\to  pD^{(*)}$
\begin{eqnarray}\label{strong2}
  {\cal M}(\Lambda_c(2595)\to  pD)&=&ig_{c1}\bar{u}_pu_{\Lambda_c}\,,\nonumber\\
  {\cal M}(\Lambda_c(2595)\to  pD^*)&=&ig_{c2}\bar{u}_p\gamma_\mu\gamma_5 u_{\Lambda_c}\epsilon^{*}_{\mu}\,.
\end{eqnarray}
Using $g_{c1}=\frac{3.69}{\sqrt{2}},g_{c2}=\frac{5.7}{\sqrt{2}}$ quoted from
Ref.~\cite{Garcia-Recio:2008rjt}~(Refs.~\cite{Hofmann:2005sw, Romanets:2012hm}
also provide similar values) and ignoring the mass difference between $\Lambda_c$
and $\Lambda_c(2595)$, one can estimate the contribution of $\Lambda_c(2595)$ by
\begin{eqnarray}\label{srt}
  &&\frac{{\cal B}(\Lambda_b\to Z^-_{cs}p)\text{ through }\Lambda_c(2595)}{ {\cal B}(\Lambda_b\to Z^-_{cs}p)\text{ through }\Lambda_c}\nonumber\\
  &\sim&\frac{\mid g_{c1}\mid^2{\cal B}(\Lambda_b\to \Lambda_c D_s^{-})+\mid g_{c2}\mid^2{\cal B}(\Lambda_b\to \Lambda_c D_s^{-})}{\mid g_{\Lambda_cpD}\mid^2{\cal B}(\Lambda_b\to \Lambda_c D_s^{-})+\mid g^V-g^T\mid^2{\cal B}(\Lambda_b\to \Lambda_c D_s^{-})}\nonumber\\
  &\leq&\frac{(\frac{3.69}{\sqrt{2}})^2\times1.72+(\frac{5.7}{\sqrt{2}})^2\times2.28}{(10.7)^2\times11.92+(-9.4)^2\times17.49}\sim(10^{-2})\,.
\end{eqnarray}
This indicates the contribution of $\Lambda_c(2595)$ is orders of magnitude
smaller than that of the ground sate $\Lambda_c$. Furthermore, using the
branching fractions and couplings given in
Refs~\cite{Liang:2016ydj,Liang:2014kra} will lead to the same conclusion.
Note that the information about ${\cal M}(\Lambda_c(2595)\to  pD^*)$
and ${\cal B}(\Lambda_b\to \Lambda_c(2595) D_s^{*-})$ is absent in
Refs~\cite{Liang:2016ydj,Liang:2014kra}, but one could assume that them are of
the same order of magnitude as ${\cal M}(\Lambda_c(2595)\to pD)$ and
${\cal B}(\Lambda_b\to \Lambda_c(2595) D_s^{-})$.

%------------------------------------------------------------------------------

\section{Conclusions}
In conclusion, the study of the hidden-charm states $Z_{cs}$ is crucial in
understanding its exotic nature. By employing different channels to produce
$Z_{cs}$, we can identify and confirm the molecular model or other models
proposed for them. Our study has focused on the triangle-rescattering decays
$\Lambda_b\to \Lambda_c D_s^{(*)-}\to Z^-_{cs}p$, where $\Lambda_c$ and
$D_s^{-}(D_s^{*-})$ transform into $Z^-_{cs}$ and $p$ by exchanging a
$D^{*}(D)$ in the triangle loop. We have proposed $\Lambda_b\to Z^-_{cs}p$
with $Z^-_{cs}\to J/\psi K^{(*)-}p$ as candidate decays to search for the
$Z_{cs}$ exotic states, and predicted
${\cal B}(\Lambda_b\to Z^-_{cs}p)=(3.1^{+1.4}_{-2.6})\times 10^{-4}$ in the
molecular model. It is worth noting that Ref.~\cite{Wan:2020oxt} also provides
the coupling values $g_{ZD_s D^*}$ and $g_{ZD_s^{*} D}$ under the assumption
that $Z_{cs}$ is a tetraquark. These couplings are approximately one third of
those in the molecular model. Consequently, the branching ratio of
$Z_{cs}\to Z^-_{cs} p$ based on the tetraquark model is an order of magnitude
smaller than that of the molecular model.

%------------------------------------------------------------------------------
We are optimistic about the prospects of our proposal being tested by the LHCb
Collaboration. With 3~fb$^{-1}$ of $pp$ collision data taken at $\sqrt{s}=7$
and 8~TeV and 6~fb$^{-1}$ at $\sqrt{s}=13$~TeV, LHCb can achieve a sensitivity
level of $10^{-6}$ for branching fractions of the $\Lambda_b$ decays.
Explicitly, the LHCb experiment has discovered exotic structures on the
$J/\psi p$ spectrum of the $\Lambda_b\to J/\psi K^-p$
events~\cite{LHCb:2019kea, LHCb:2015yax, LHCb:2015qvk}. Our proposal is based
on the same final state particles, while on the $J/\psi K^-$ spectrum,
indicating an accessible opportunity to search for the $Z_{cs}$ exotic states
via the $\Lambda_b\to Z^-_{cs}p$ decay.

\section*{ACKNOWLEDGMENTS}
The authors thank Dr.~Chia-Wei Liu for helpful discussions. YY was supported in part by National Natural Science Foundation of China (NSFC) under Contracts No.~11905023, No.~12047564 and No.~12147102, the Natural Science Foundation of Chongqing (CQCSTC) under Contract No.~cstc2020jcyj-msxmX0555 and the Science and Technology Research Program of Chongqing Municipal Education Commission (STRPCMEC) under Contracts No.~KJQN202200605 and No.~KJQN202200621; HZ and BCK were supported in part by NSFC under Contracts No.~11875054 and No.~12192263 and Joint Large-Scale Scientific Facility Fund of the NSFC and the Chinese Academy of Sciences under Contract No.~U2032104. JWZ was supported by NSFC under Contract No.~12275036, CQCSTC under Contract No.~cstc2021jcyj-msxmX0681 and STRPCMEC under Contract No.~KJQN202001541. DZH is supported by NSFC under Contract No.~12275037 and STRPCMEC under Contract No.~KJQN202300609. RYZ was supported by CQCSTC under Contract No.~CSTB2022NSCQ-MSX0534, and STRPCMEC under Contract No.~KJQN202300614.

\begin{appendices}
%------------------------------------------------------------------------------
\section{multi-point function}\label{point}
The integrations of the multi-point functions are given
by~\cite{tHooft:1978jhc,Passarino:1978jh,Hahn:1998yk,Denner:2005nn}
\begin{eqnarray}\label{int}
  &&\{C_{0};C_1^{\mu};C_2^{\mu\nu};C_3^{\mu\nu\alpha}\}\,\nonumber\\
  &&=\int\frac{d^{4}q_2}{i \pi^2}\frac{\{1;q_2^{\mu};q_2^{\mu}q_2^{\nu};q_2^{\mu}q_2^{\nu}q_2^{\alpha}\}}
  {(q_2^{2}-m_{2}^{2}+i\epsilon)[(q_2-p_1+p_3)^{2}-m^{2}_{1}+i\epsilon][(q_2+p_3)^{2}-m_{3}^{2}+i\epsilon]}\,,\nonumber\\
  &&\{D_2^{\mu\nu};D_3^{\mu\nu\alpha}\,;D_4^{\mu\nu\alpha\beta}\}\nonumber\\
  &&=
  \int\frac{d^{4}q_2}{i \pi^2}\frac{\{q_2^{\mu}q_2^{\nu};q_2^{\mu}q_2^{\nu}q_2^{\alpha};q_2^{\mu}q_2^{\nu}q_2^{\alpha}q_2^{\beta}\}}
           {(q_2^{2}+i\epsilon)(q_2^{2}-m_{2}^{2}+i\epsilon)[(q_2-p_1+p_3)^{2}-m^{2}_{1}+i\epsilon][(q_2+p_3)^{2}-m_{3}^{2}+i\epsilon]}\,.
\end{eqnarray}
The analytical results of the scalar point functions, $B_{0}$, $C_{0}$, and
$D_{0}$, can be found in Refs.~\cite{tHooft:1978jhc}.
All other point functions can be expressed as the linear combination of scalar
point functions:
%------------------------------------------------------------------------------
\begin{eqnarray}\label{cij_1}
  C_1^{\mu}&=&p_1^\mu C_{11} + p_3^\mu C_{12}\,,\nonumber\\
  C_2^{\mu\nu}
  &=&p_1^\mu p_1^\nu C_{21}
  +p_3^\mu p_3^\nu C_{22}
  +(p_1^\mu p_3^\nu+p_1^\nu p_3^\mu)C_{23}+g^{\mu\nu}C_{24}\,,\nonumber\\
  D_2^{\mu\nu}
  &=&p_1^\mu p_1^\nu D_{21}
  +p_3^\mu p_3^\nu D_{22}
  +(p_1^\mu p_3^\nu+p_1^\nu p_3^\mu)D_{23}+g^{\mu\nu}D_{24}\,,\nonumber\\
  D_3^{\mu\nu\alpha}
  &=&p_1^\mu p_1^\nu p_1^\alpha D_{31}+(p_1^\mu p_3^\nu p_3^\alpha+p_3^\mu p_1^\nu p_3^\alpha+p_3^\mu p_3^\nu p_1^\alpha) D_{33}\nonumber\\
  &&+p_3^\mu p_3^\nu p_3^\alpha D_{32}+(p_1^\mu p_1^\nu p_3^\alpha+p_3^\mu p_1^\nu p_1^\alpha+p_1^\mu p_3^\nu p_1^\alpha) D_{34}\nonumber\\
  &&+(p_1^\mu g^{\nu\alpha}+p_1^\nu g^{\mu\alpha}+p_1^\alpha g^{\mu\nu}) D_{35}+(p_3^\mu g^{\nu\alpha}+p_3^\nu g^{\mu\alpha}+p_3^\alpha g^{\mu\nu}) D_{36}\,,\nonumber\\
  D_4^{\mu\nu\alpha\beta}
  &=&p_1^\mu p_1^\nu p_1^\alpha p_1^\beta D_{41}+(p_3^\mu p_1^\nu p_1^\alpha p_1^\beta+p_1^\mu p_3^\nu p_1^\alpha p_1^\beta+p_1^\mu p_1^\nu p_3^\alpha p_1^\beta+p_1^\mu p_1^\nu p_1^\alpha p_3^\beta) D_{43}\nonumber\\
  &&+p_3^\mu p_3^\nu p_3^\alpha p_3^\beta D_{42}+(p_1^\mu p_3^\nu p_3^\alpha p_3^\beta+p_3^\mu p_1^\nu p_3^\alpha p_3^\beta+p_3^\mu p_3^\nu p_1^\alpha p_3^\beta+p_3^\mu p_3^\nu p_3^\alpha p_1^\beta) D_{44}\nonumber\\
  &&+(p_1^\mu p_1^\nu p_3^\alpha p_3^\beta+p_1^\mu p_3^\nu p_1^\alpha p_3^\beta+p_1^\mu p_3^\nu p_3^\alpha p_1^\beta+p_3^\mu p_1^\nu p_1^\alpha p_3^\beta+p_3^\mu p_1^\nu p_3^\alpha p_1^\beta+p_3^\mu p_3^\nu p_1^\alpha p_1^\beta) D_{44}\nonumber\\
  &&+(p_1^\mu p_1^\nu g^{\alpha\beta}+p_1^\mu p_1^\alpha g^{\nu\beta}+p_1^\mu p_1^\beta g^{\nu\alpha}+p_1^\nu p_1^\alpha g^{\mu\beta}+p_1^\nu p_1^\beta g^{\mu\alpha}+p_1^\alpha p_1^\beta g^{\mu\nu}) D_{46}\nonumber\\
  &&+(p_3^\mu p_3^\nu g^{\alpha\beta}+p_3^\mu p_3^\alpha g^{\nu\beta}+p_3^\mu p_3^\beta g^{\nu\alpha}+p_3^\nu p_3^\alpha g^{\mu\beta}+p_3^\nu p_3^\beta g^{\mu\alpha}+p_3^\alpha p_3^\beta g^{\mu\nu}) D_{47}\nonumber\\
  &&+(p_1^\mu p_3^\nu g^{\alpha\beta}+p_1^\mu p_3^\alpha g^{\nu\beta}+p_1^\mu p_3^\beta g^{\nu\alpha}+p_1^\nu p_3^\alpha g^{\mu\beta}+p_1^\nu p_3^\beta g^{\mu\alpha}+p_1^\alpha p_3^\beta g^{\mu\nu}\nonumber\\
  &&+p_3^\mu p_1^\nu g^{\alpha\beta}+p_3^\mu p_1^\alpha g^{\nu\beta}+p_3^\mu p_1^\beta g^{\nu\alpha}+p_3^\nu p_1^\alpha g^{\mu\beta}+p_3^\nu p_1^\beta g^{\mu\alpha}+p_3^\alpha p_1^\beta g^{\mu\nu}) D_{48}\nonumber\\
  &&+(g^{\mu\nu} g^{\alpha\beta}+g^{\mu\alpha} g^{\nu\beta}+g^{\mu\beta} g^{\nu\alpha}) D_{49}\,,
\end{eqnarray}
where $C_{ij}$ and $D_{ij}$ are the linear combination of $B_{0}$, $C_{0}$, and
$D_{0}$ defined in
Refs~\cite{Passarino:1978jh,Hahn:1998yk,Denner:2005nn}.
\end{appendices}
\end{document}